\documentclass[prl,twocolumn,aps,amsmath,amssymb]{revtex4-1}
\usepackage{graphicx}
\usepackage{dcolumn}
\usepackage{bm}
\usepackage{hyperref}
\usepackage{color}
\usepackage{times}
\usepackage{amsmath, amsthm, amssymb}
\usepackage{subfigure}
\usepackage{setspace}
\usepackage{bm}

\begin{document}

\title{Entanglement and spin-squeezing in a network of optical lattice clocks}

\author{Eugene S. Polzik$^1$ and Jun Ye$^2$}
\affiliation{
$^1$Niels Bohr Institute, University of Copenhagen; Blegdamsvej 17, 2100 Copenhagen, Denmark\\
$^2$JILA, National Institute of Standards and Technology and University of Colorado, Boulder, CO 80309-0440, USA
}

\begin{abstract}
We propose an approach for collective enhancement of precision for remote optical lattice clocks and a way of generation of the Einstein-Podolsky-Rosen (EPR) state of remote clocks.  In the first scenario a distributed spin squeezed state (SSS) of $M$ clocks is generated by a collective optical quantum nondemolition measurement on clocks with parallel Bloch vectors. Surprisingly, optical losses which usually present the main limitation to SSS can be overcome by an optimal network design which provides close to Heisenberg scaling of the time precision with the number of clocks $M$. We provide an optimal network solution for distant clocks as well as for clocks positioned in close proximity of each other. In the second scenario, we employ collective dissipation to drive two clocks with oppositely oriented Bloch vectors into a steady state entanglement. The corresponding EPR state provides secret time sharing beyond the projection noise limit between the two quantum synchronized clocks protected from eavesdropping. An important application of the EPR entangled clock pair is remote sensing of, for example, gravitational effects and other disturbances to which clock synchronization is sensitive. \\PACS {32.80.Wr, 37.30.+i, 42.50.Ct, 42.62.Fi}
\end{abstract}

\date\today
\maketitle

\noindent\textbf{Introduction.} Optical atomic clocks provide some of the most precise and accurate physical measurements to date~\cite{Bloom2014,Nicholson2015,Katori2015,Chou2010,Madej2012}. The precision of optical lattice clocks is presently limited by the available frequency stability of the best lasers~\cite{Hinkley2013,Bishof2013,Kessler2012}, but quantum noise of uncorrelated atoms lures not far below. With enhanced laser stability and improved measurement protocol to reduce the laser noise~\cite{Borregaard2013,Rosenband2013,Kohlhaas2015}, the next frontier of precision can be advanced by generating entangled states of the clock atoms. For small number of atoms $N$, a maximally entangled clock operating with GHZ states can reach the Heisenberg limit of stability much faster than the best classical schemes~\cite{Kessler2014,Komar2014}. Spin-squeezed states (SSS)~\cite{wineland1992} are particularly suitable for improving the precision of optical lattice clocks that operate on large $N$ and currently hold the record for clock precision~\cite{Nicholson2015}. Distant clocks connected into a spin squeezed network can provide a higher collective precision for all users.

Spin squeezing (SS) and entanglement of atomic ensembles have so far been experimentally demonstrated for single ensembles of spins associated with atomic states separated by radio- or microwave frequencies. This was achieved by optical quantum nondemolition (QND) measurement~\cite{hald1999,kuzmich2000, julsgaard2001, appel2009,schleier2010, Leroux2010, takano2009, thompson2014}, by mapping squeezed light onto an atomic ensemble~\cite{appel2008}, by atomic interactions in a Bose-Einstein condensate~\cite{esteve2008}, and by engineered dissipation~\cite{krauter2011}. Improvement to clock precision beyond the quantum projection noise (QPN) was demonstrated for microwave clocks~\cite{appel2009,Leroux2010}.

Networks of remote clocks offer new possibilities for secret time sharing, remote sensing and interferometry that can take advantage of the unprecedented clock precision. A recent proposal outlined probabilistic generation of GHZ-type of entanglement by single photon communication between distant clocks each containing a small number of $n_q$ qubits as discrete quantum variables~\cite{Kessler2014,Komar2014}. However, for optical lattice clocks containing macroscopic numbers of atoms $N$ we encounter continuous variables such as spin squeezed and EPR entangled states that can be generated deterministically. QND probing on cyclic transitions has been identified as the condition for Heisenberg scaling with $N$ in Ref.~\cite{Saffman2009} and demonstrated for microwave clocks in Ref.~\cite{appel2009,thompson2014}. However, Heisenberg scaling with the number of clocks $M$ in a chain is problematic as SSS have low tolerance for losses, in particular to losses of the optical channel for optical QND. Here we demonstrate that an optimal design of the network of cavity-enhanced optical clocks allows to keep Heisenberg scaling with $M$ even in the presence of substantial channel losses.  Optical lattice clocks with their long coherence times are ideal for generation of such states.

In the second part of the Letter we present a scenario where an Einstein-Podolsky-Rosen (EPR) entangled state of two clocks is generated by engineered dissipation. Such clocks feature the ''synchronized time'' protected from any eavesdropper and available only for the participants working together.
Both proposals are aimed at optical clocks with a macroscopic number of atoms. As a specific example, we show their feasibility for Sr clocks with realistic experimental parameters.

\noindent\textbf{A network of clocks in a collective squeezed state.}

The ultimate limit of precision for a clock made of $N$ independent atoms is defined by the the angular uncertainty of a coherent spin state (CSS) of the ensemble (quasi-)spin~\cite{itano1993}. CSS is a product state $|\Psi\rangle = \Pi_{i=1}^{N} \frac{1}{\sqrt{2}}
(|1\rangle_i+|2\rangle_i)$ of uncorrelated atoms oriented in the same direction, $J_x=N/2$. The other two projections of $\hat{J}$ have minimal equal variances allowed by the
Heisenberg uncertainty relation: $Var(J_z)=Var(J_y) = J_x/2 = N/4$. Introducing quantum correlations between atoms allows to reduce $Var(J_z)$ to below the QPN limit. For the resulting spin squeezed state (SSS), under the condition that
\begin{align}
  \xi = {}&
  \frac{Var(J_z)}{J^2}N =Var(X_A)\frac{N}{J}<1
  \label{eq:wineland},
\end{align}
the atoms become entangled~\cite{sorensenduan2001}, and the corresponding signal-to-noise ratio for spectroscopy is improved by the inverse of $\xi$ ~\cite{wineland1992}, which is the spin squeezing parameter for metrology. The canonical operators, $X_A=J_z/\sqrt{J},P_A=J_y/\sqrt{J}$  obey the commutation relation $[X_A,P_A]=i$  where $J$ is the length of the mean pseudo-spin vector. The quantum noise limited clock precision defined as the minimal detectable angle of spin rotation for the clock sequence is $\sqrt{Var(J_z)}/J=\sqrt{\xi/N}$.

As shown theoretically~\cite{Saffman2009} and experimentally~\cite{appel2009} for Cs clocks, QND probing of the clock levels using cyclic transitions under a high optical depth leads to Heisenberg scaling of precision with $N$. This approach led to the recent demonstration of SSS for Rb ground state~\cite{thompson2014}.

The QND interaction $H \propto X_A X_L$ ~\cite{RMP} leads to the input-output relation for photonic canonical variables $X_L$ and $P_L$
\begin{equation}\label{eq:QND}
 P^{out}_L=P^{in}_L+\kappa X_A.
\end{equation}
with $\kappa = \sqrt{d\eta e^{-\eta}}$ the interaction constant, $d$ the resonant single pass optical depth, $\eta=n_{dr}(\gamma/\Delta)^2\sigma/A$  a parameter describing spontaneous emission caused by the probe,  $\sigma$  the resonant dipole cross section, $A$  the beam cross section, $\gamma,\Delta$ the natural linewidth and detuning of the optical transition, $n_{dr}$  the photon number in the QND probe. For QND on a cyclic transition the degree of spin squeezing is
\begin{equation} \xi=\frac{1}{e^{-\eta}\, (1+\kappa^2)}.
\label{eq:etaeqn}
\end{equation}
This results in $\xi_{min}=\sqrt{e}/{\kappa_{opt}^2}= 2e/d\propto N^{-1}$  achieved for $\eta=1/2$ and the Heisenberg scaling of the clock  precision $\sqrt{\xi/N}$ with the atom number. Such scaling for a microwave clock probed on two cyclic transitions has been demonstrated in ~\cite{appel2009,anne}.

The unique energy level structure of alkaline earth atoms provides an ideal configuration to implement a QND protocol based on a cyclic transition now in an optical clock.  As a specific example, we consider an optical lattice clock operating on the $^1$S$_0$ ($|1\rangle$) - $^3$P$_0$ ($|2\rangle$) transition with $N$ atoms placed in an optical resonator.  The collective QND readout is performed on the cyclic $^1$S$_0$ ($|1\rangle$) - $^3$P$_1$ ($|3\rangle$) narrow transition ($\sim$7.4 kHz) in Sr using a far detuned probe (Fig.~\ref{Fig:level}(b)). The clock sequence and the details of the QND measurement are outlined in the Supplementary Material.

\begin{figure}
\center\includegraphics[width=1\columnwidth]{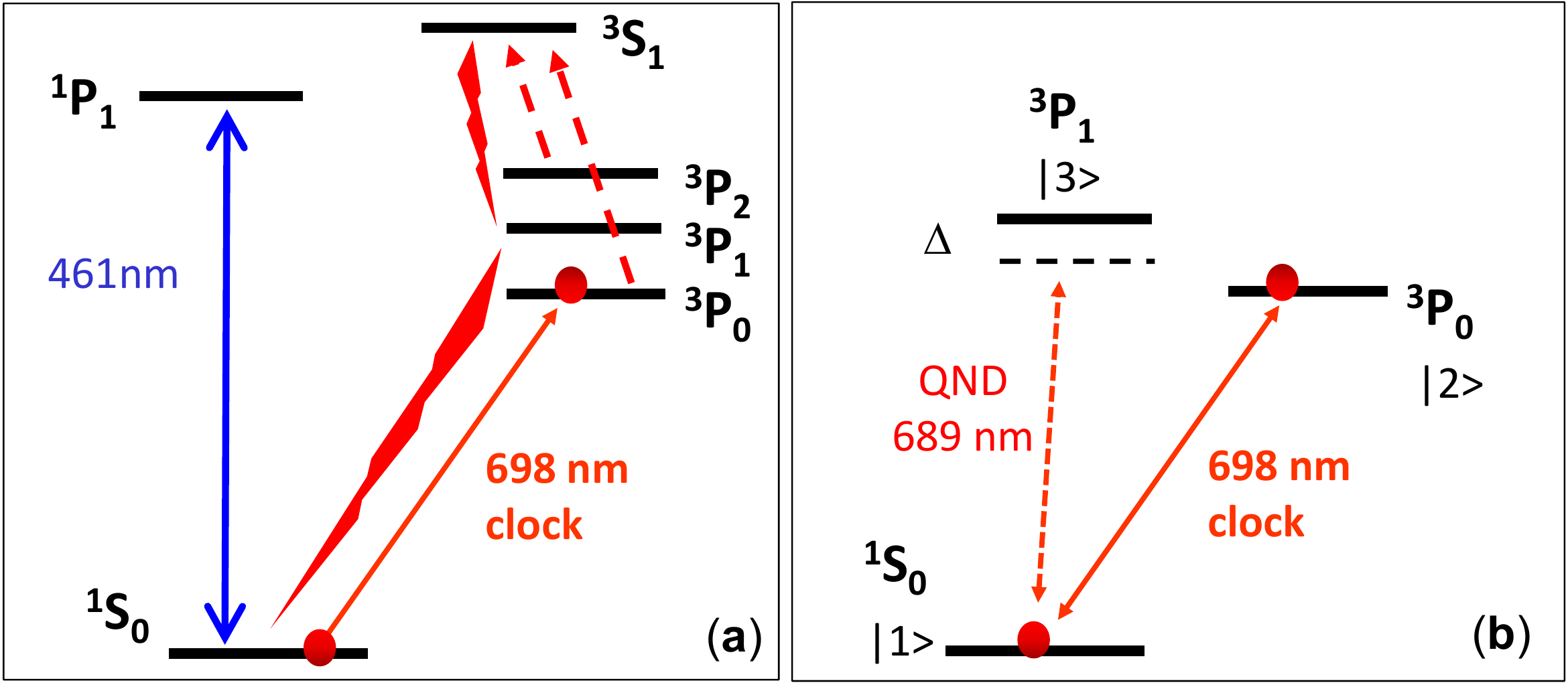}
\caption{\textbf{Clock operation and probe}. The atomic level structure for the optical lattice clock. (a) Traditional destructive readout of the clock state populations in $|1\rangle$ and $|2\rangle$. (b) QND probe of the clock transition $|1\rangle - |2\rangle$ using a far detuned probe on $|1\rangle - |3\rangle$. The wavelengths are given for Sr.}\label{Fig:level}
\end{figure}



\begin{figure}
\center\includegraphics[width=1\columnwidth]{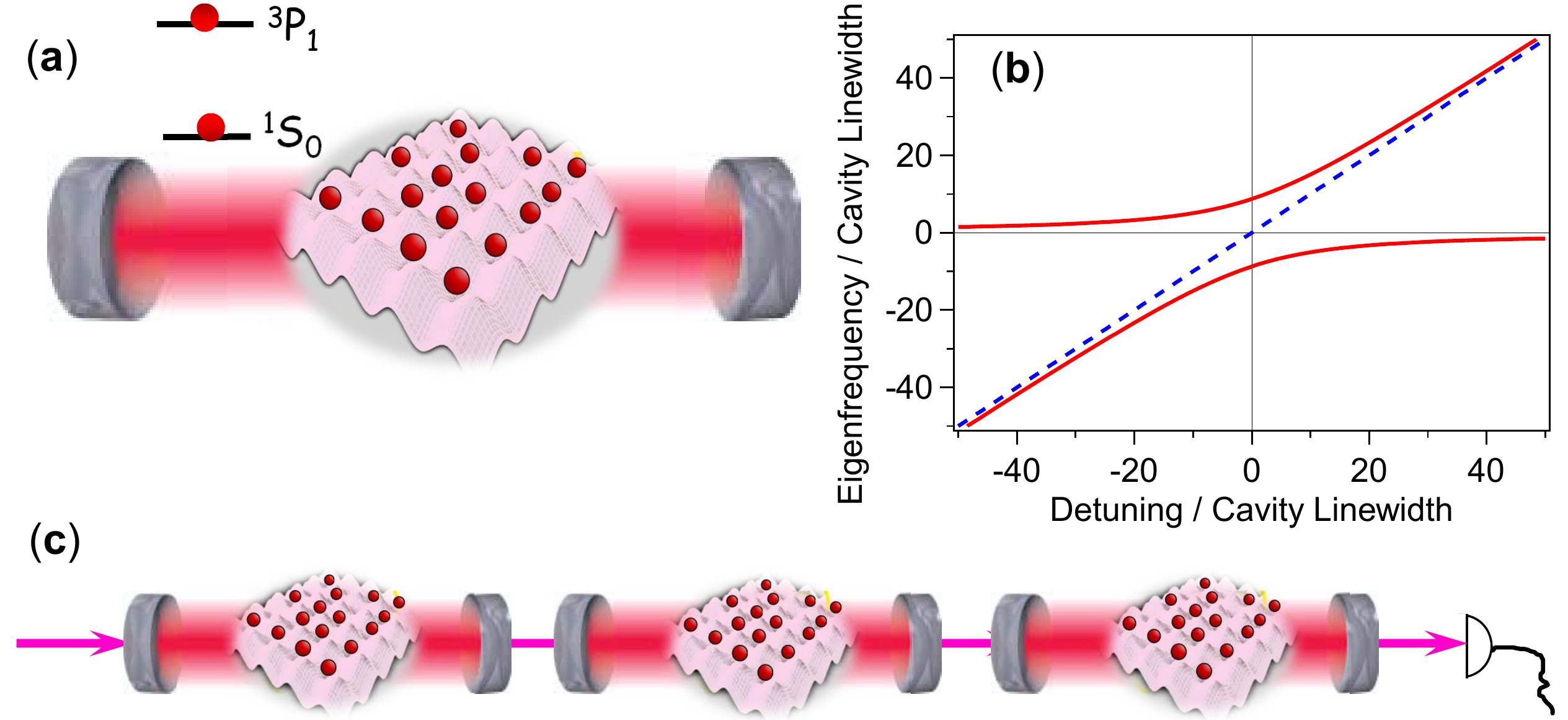}
\caption{\textbf{Entangled clock sequence} (a) Cavity QED used for QND probe of the clock states. (b) The eigenfrequency spectrum for $10^3$ Sr atoms distributed in an optical lattice inside a 5 cm cavity with $F=10^5$. The single-atom effective vacuum Rabi frequency is 16 kHz. (c) A cascaded cavity system to entangle multiple independent spin-squeezed clocks.}\label{Fig:cavity}
\end{figure}

Towards our aim of demonstrating Heisenberg scaling with the number of distant clocks in a clock chain we consider first a single lattice clock placed in an optical resonator (Fig.~\ref{Fig:cavity}(a))~\cite{Chen2014}.  In the following we assume that the detuning of the atom and cavity resonances is much greater than the vacuum Rabi frequency $\Omega$, which in turn is much greater than the atomic ($\gamma$) and the cavity ($\Gamma$) linewidths.

We consider a standing wave cavity with input/output mirror power transmission coefficients $T_1,T_2$, single pass intracavity losses $\mathcal{L}$ and the detuned probe single pass absorption  $d_{\Delta}$. The cavity power transmission coefficient on resonance for small  $\mathcal{L},d_{\Delta}$  is $4T_1T_2/(T_1+T_2+ 2\mathcal{L}+2d_{\Delta})^2=4T_1T_2/(T_1+T_2+ 2\mathcal{L})^2[1-2Fd_{\Delta}/\pi]$, where $F=2\pi /(T_1+T_2+ 2\mathcal{L})$ is the cavity finesse. Thus, $d_{\Delta}$, as well as the corresponding phase shift, is enhanced by a factor $2F/\pi$. Depending on details of the experimental realization, the optimal measurement is achieved either in reflection from a single-ended over-coupled cavity with $T_1 \gg T_2, \mathcal{L}, d_{\Delta}$ or with a symmetric cavity in transmission. The atomic absorption, $d_{\Delta}=N/{2n}$ at the optimal $\eta=1/2$, can be reduced by using larger $\Delta$ and photon number $n$.

Eq.~\ref{eq:QND} is modified in the presence of the cavity. With $n$ photons detected during the interaction time the observed probe phase shift consists of two terms:
\begin{equation}\label{eq:phaseshift}
\varphi = n^{-1/2}+  \sqrt{d e^{-\eta_n}} (\gamma/\Delta)\sqrt{\sigma/A}2F/{\pi}X_A.
\end{equation}
The first term is the shot noise of detected light, and the second term represents the cavity enhanced phase shift. To derive the cavity-based input-output equation we multiply both sides of Eq.\ref{eq:phaseshift} with $\sqrt{n}$,
\begin{align}\label{eq:cavityenhanced}
 P^{out}_L = {}& P^{in}_L+\kappa_{cav}X_A  \nonumber\\
 = {}& 1 + \sqrt{d n e^{-\eta_n}} (\gamma/\Delta)\sqrt{\sigma/A}2F/{\pi}X_A.
\end{align}
Here, $\kappa_{cav}=\sqrt{d n e^{-\eta}} (\gamma/\Delta)\sqrt{\sigma/A}2F/{\pi}=\sqrt{d \eta_n e^{-\eta_n} }2F/{\pi}$ is the cavity enhanced atom-light interaction constant and $\eta_n$ corresponds to the detected photon number $n$. The relation between the spontaneous emission rate in the cavity and in free space is $\eta_{cav}=\eta_n F/\pi$ 
for small $\mathcal{L}$. Eq.~\ref{eq:etaeqn} is then modified with substitutions $\eta \rightarrow \eta_{cav}=1/2$ and $\kappa_{opt} \rightarrow \kappa_{cav}=\sqrt{4dF\eta_{cav}e^{-\eta_{cav}}/\pi}=\sqrt{2dFe^{-1/2}/\pi}$ with the optimal value $\eta=1/2$. For the case of large optical depth and/or finesse, $4d \eta_n  F^2/{\pi}^2\gg 1$, we arrive at the squeezed spin variance $\xi_{min}=\sqrt{ e}/{\kappa_{cav}^2}= e\pi/(2dF)\propto (FN)^{-1}$ achieved for $\eta_{cav}=1/2$, valid for $dF\gg 1$. The clock precision is then $\sqrt{\xi_{min}/N}=\sqrt{2\pi eA/(\sigma F)}N^{-1}$. Note that our treatment is limited to $ \sigma F/A<2\pi e$, otherwise the Holstein-Primakoff approximation breaks down when the size of the antisqueezed component becomes comparable with $N$.

In a realistic design for cavity QED with Sr atoms~\cite{Bishof2014}, we envision $F=10^5$, length of 5 cm, and $N$ = 1000. The atoms in state $|1\rangle$ are collectively coupled to a single mode of this cavity through state $|3\rangle$. The bare cavity mode is dressed by the presence of the $|1\rangle$-atoms and the resonance spectrum is determined by the atom-cavity detuning ($\Delta$) and the collective vacuum Rabi splitting ($\Omega$) that depends on the number of atoms in $|1\rangle$ (Fig.~\ref{Fig:cavity}(b)). The estimated $\Omega=500$ kHz, which is to be compared with $\Gamma=29$ kHz and $\gamma=7$ kHz. The collective cooperativity factor $\Omega^2/(\Gamma\gamma)=d\,F$ = 1200, leading to an estimated 20 dB of metrologically useful spin squeezing. For Sr with a cyclic optical transition, the actual value of $\Delta$ does not play a fundamental role, but $\Delta\gg\Omega$ can be useful if large values of $n$ are desired.


It follows from the above discussion that in the case of a lossless optical channel connecting $M$ identical clocks, a collective QND probe of the whole network leads to the precision that is a factor of $M$ better than each clock, as opposed to uncorrelated clocks for which the precision improves by a factor of $\sqrt{M}$. Figure~\ref{Fig:cavity}(c) shows an optical probe field passing through a chain of successive optical cavities followed by a single quantum measurement performed at the output. Such interaction generates a collective squeezed state of the entire system of $M$ clocks. A channel with finite losses can be accounted for by the substitution  $\kappa =\sqrt{4dF\eta e^{-\eta}/\pi} \rightarrow \kappa_i=\sqrt{4dF_i\eta_i e^{-\eta_i }e^{-r_i}/\pi}$, with the probe induced decoherence for the $i$-th clock $\eta_i$,
where $e^{-r_i}$ describes the optical channel transmission from the $i$-th clock to the detector (subscript $cav$ omitted for brevity). The noise of the measurement is still the shot noise of the detected probe whereas the signals due to the spin projection from all clocks add up, so that $S/N$ for the chain is $ \sum\limits_{i=1}^M \sqrt{N}\kappa_i e^{-\eta_i/2}=\sum\limits_{i=1}^M \sqrt{4dNF_i\eta_i e^{-2\eta_i}e^{-r_i}/\pi }$. Maximal collective spin squeezing for the chain can be found by optimizing this expression, given the clock parameters and the channel transmission properties. Consider, for example, $M$ clocks connected with a channel with equal transmission $e^{-r}$ between each pair of clocks ($r_i=(M-i)r$ and total channel transmission is $t=e^{-(M-1)r}$). With the optimal value $\eta_i=ne^{r_i}(\gamma/\Delta)^2\sigma/AF_i/\pi=1/2$, the collective S/N becomes $\sqrt{4dNF_M /\pi e}\sum\limits_{i=1}^M e^{-(M-i)r}= \sqrt{4dNF_M /\pi e}(e^{-Mr}-1)/(e^{-r}-1)$ where the fixed value of $\eta_i$ dictates that the cavity finesse is maximal for the last clock in the chain, $F_i=F_M e^{(i-M)r}$.

Assuming a sufficiently dense chain of distant clocks ($r$$\ll$1 but $Mr$$\gg$1), we reach the precision for the chain $(S/N)^{-1}$ = $(4dNF_M /\pi e)^{-1/2}M^{-1}$$\mid$$\ln{t}$$\mid$$ \propto (NM)^{-1}$$\mid$$\ln{t}$$\mid$. The expression in parenthesis is limited to $\ll$$N$ because the size of the antisqueezed quadrature must be $\ll$$N$.  Within this limit we obtain Heisenberg scaling of the precision of the chain with both $N$ and $M$ for any given channel transmission $t$. For example, four clocks probed by QND measurement through a channel with $t=e^{-(M-1)r}=0.5$ ($3$ dB total losses) provide precision improvement of $3.1$, and eight clocks in the same channel give the improvement by $6$. If the ultimate performance of each clock dictates an upper limit on $N$ due to, $e.g.$, atomic interactions~\cite{Nicholson2012}, a chain of entangled clocks may provide an optimal solution. Distant clocks in a collective SSS may offer an opportunity for testing sensitive relativistic effects~\cite{brukner}.



\begin{figure}
\center\includegraphics[width=1\columnwidth]{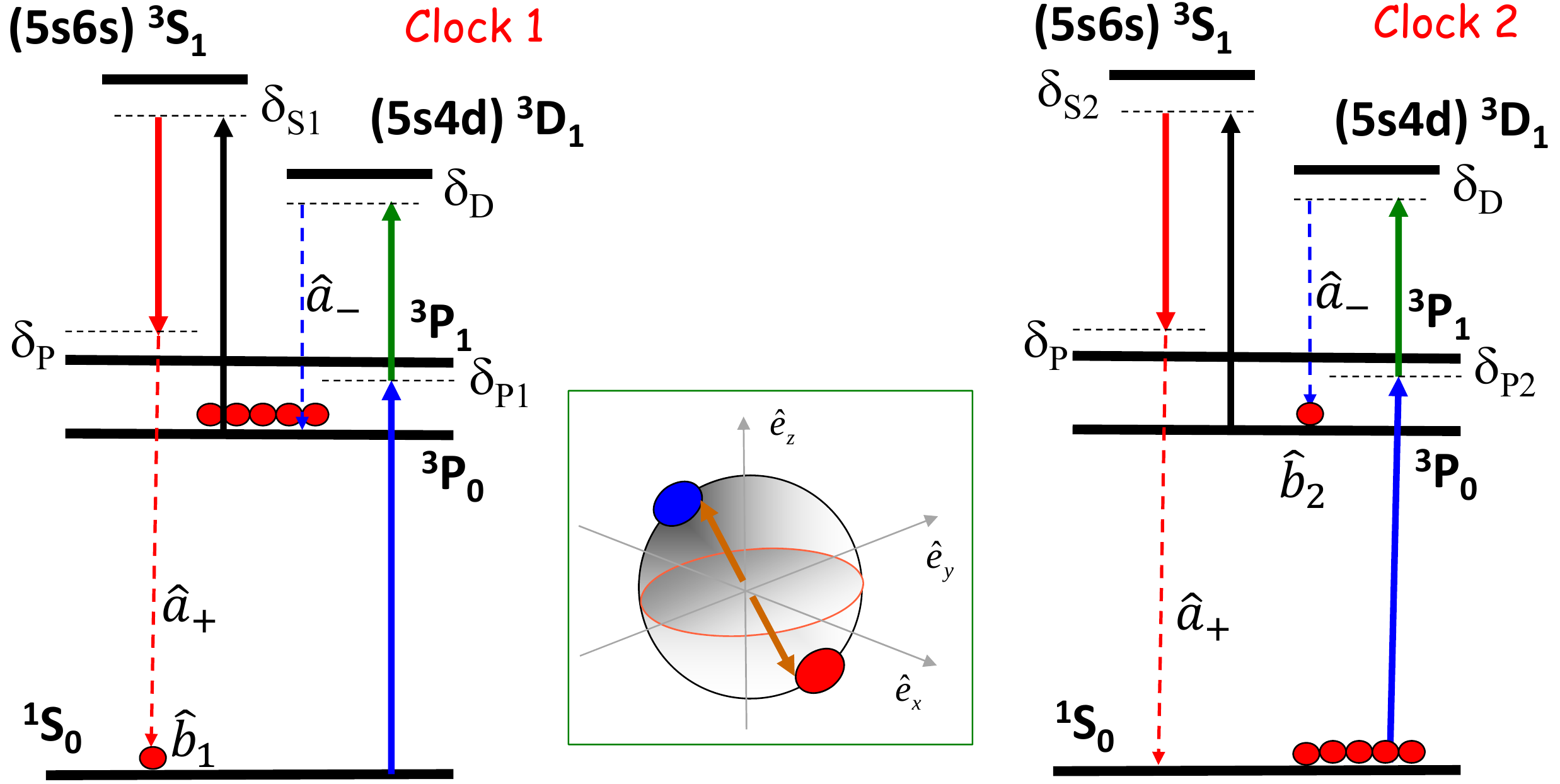}
\caption{\textbf{Transitions driving two clocks into an EPR-entangled pair.}  Clock 1 and clock 2 are driven with four phase locked classical fields (solid arrows). Forward scattered quantum fields (dashed arrows) generate entanglement corresponding to the two clock Bloch vectors being exactly antiparallel despite their individual quantum noise.}
\label{Fig:EPRlevels}
\end{figure}

\noindent\textbf{EPR entangled clocks.} SSS discussed above allows for determination of one of the two quantum projections of the Bloch vector to better than $\sqrt{J/2}$, which is sufficient for improved clock precision. However, for a pair of suitably designed clocks a more intriguing state is possible where \textit{both} projections are defined better than this limit \textit{with respect to each other}. Clock comparison can thus run significantly better than the conventional synchronous mode~\cite{Bize2000,Takamoto2011,Nicholson2012}. Such state of two Bloch vectors (spins) is a special case of the EPR state with the entanglement condition $Var(J_{y1}+J_{y2})+ Var(J_{z1}+J_{z2})<2J$ ~\cite{RMP}. It can be realized when the mean spins of the two ensembles are oriented in opposite directions, $J=J_{x1}=-J_{x2}$,  as demonstrated for collective magnetic spins~\cite{julsgaard}. For optical clocks the requirement of oppositely oriented mean spins means that the two clocks should be initialized in two opposite clock states (Fig. ~\ref{Fig:EPRlevels}).

When the Bloch vector describes a pair of states separated by an optical transition, the conventional QND method of generating an EPR state is not applicable because it would require a direct measurement of the oscillations at an optical frequency. However, as demonstrated for magnetic spin oscillators~\cite{wasilewski2009}, the EPR state can be generated by a common dissipation process provided by forward scattering of indistinguishable photons that does not involve any measurement. The interaction Hamiltonian between two atomic ensembles and light that generates an EPR state of the atomic operators $b^{\dagger}_1$ and $b^{\dagger}_2$ is $H \propto \mu_1 a^{\dagger}_{+} b^{\dagger}_1+ \nu_1 a^{\dagger}_{-} b_1 + \mu_2 a^{\dagger}_{-} b^{\dagger}_2 + \nu_2 a^{\dagger}_{+} b_2 + h.c.$  The first (last) two terms describe the creation of photon fields $a_+,a_-$ and corresponding creation/annihilation of the collective atomic excitation $b_1$ ($b_2$). For an optical clock these operators correspond to the collective excitation generated in the lower (upper) state of clock 1 (2). Entanglement is generated under the following conditions ~\cite{wasilewski2009}: photons scattered from the two clocks into mode $a_+$ are indistinguishable (same for mode $a_ -$) and $\mu_1=\mu_2, \nu_1=\nu_2$.

The challenge of realization of such interaction for an optical clock transition (or for any collective excitation scheme realized on an optical transition) is that due to the selection rules the above conditions are not feasible with a standard Raman transition (four-wave mixing). It turns out, however, that these conditions can be fulfilled using a six-wave mixing process shown for a specific example of Sr optical clock in (Fig.~\ref{Fig:EPRlevels}). The use of two-photon driving fields (blue, red, black and green solid arrows) allows to fulfill the condition of indistinguishability  for photons $a_+$ ($a_-$) emitted by the two ensembles by choosing the two-photon detunings $\delta_P,\delta_D$ to be the same in both clocks and by phase locking of the lasers (solid arrows). The condition $\mu_1=\mu_2, \nu_1=\nu_2$ for scattering amplitudes in the Hamiltonian can be met by tuning the one-photon detunings $\delta_{S1},\delta_{S2},\delta_{P1},\delta_{P2}$. Similar to SS, the degree of entanglement scales with the optical depth thus benefiting from cavity enhancement as well.

An ideal entangled state of this kind corresponds to the two clock Bloch vectors being exactly antiparallel (Fig.~\ref{Fig:EPRlevels}). This is to be contrasted with the case of SSS where the Bloch vector direction is defined better than QPN only in the plane in which the squeezing axis lies.

The EPR state can be used for secret time sharing analogous to the quantum key distribution. The clock sequence resembles the standard clock sequence (see Fig.~\ref{Fig:sequence} in Supplementary Material) with an important inset in step (d). At this step the two clock owners randomly choose either to apply or not to apply the $\pi/2$ rotation around $x$ axis.  They then publicly exchange the choice with respect to the $\pi/2$ pulses, but not the results of the clock interrogation. The procedure is repeated several times. In close analogy to the quantum key distribution we can use the measurements in which we have made the same choice of rotations for the relative time measurements with high precision. Each of the clock owners acting separately will achieve a much worse precision, compared to QPN, since one half of the EPR state is a noisy thermal state. If an eavesdropping attempt is made, the combined two-clock precision will be compromised as well.

Another attractive feature of the EPR entangled clocks is the improved capability to check any clock disagreement quickly, enabling an efficient approach for characterization of systematic effects of an unknown clock (2) using a well-calibrated clock (1).

Perhaps the most important application of an EPR pair of clocks is for remote sensing.  The EPR correlation can be used to map out electromagnetic field from site 1 to site 2 remotely.  Namely, one can slave clock 2 to clock 1 by matching the conditions of clock 1 to that of 2.  In fact, this might be the best tool to explore gravitation potential-induced decoherence such as described in Ref.~\cite{pikovski}, and it can also serve a potentially important role for a future long-baseline atom inteferometer for gravitational wave detection~\cite{Hogan2015}.

We thank A. S\o rensen, M. Bishof, X. Zhang, S. Bromley, and M. Barrett for discussions. We acknowledge funding from ERC grant INTERFACE, NIST, DARPA, and NSF-JILA PFC. We dedicate this work to H. J. Kimble, as the initial ideas were formed at a celebration of his scientific achievement in April 2014.

\newpage

\noindent\textbf{SUPPLEMENTARY MATERIAL}

\noindent\textbf{Quantum nondemolition measurement in an optical clock. } We consider an optical lattice clock operating on the $^1$S$_0$ ($|1\rangle$) - $^3$P$_0$ ($|2\rangle$) transition with $N$ atoms placed in an optical resonator. In a normal clock operation, the population of the two clock states is measured destructively by scattering photons with the strong $^1$S$_0$ - $^1$P$_1$ transition (Fig.~\ref{Fig:level}(a)), permitting a QPN-limited probe of the clock transition. A phase sensitive probe based on this strong transition was implemented to enable a less destructive measurement of the state population~\cite{Lodewyck2009}. Here we consider a collective readout of the cyclic $^1$S$_0$ ($|1\rangle$) - $^3$P$_1$ ($|3\rangle$) narrow transition ($\sim$7.4 kHz) in Sr using a far detuned probe (Fig.~\ref{Fig:level}(b)).

An ensemble of $N$ clock atoms can be described by the collective pseudo-spin vector $\hat{J}$ of spin-$1/2$ particles. $J_z$ is defined by the population
difference $\Delta N$, such that: $J_z=\frac 1 2 (N_1-N_2)=\Delta N/2$. Atoms are prepared in a superposition of the two clock states by a $\pi /2$ rotation of $\hat{J}$ around the $y$-axis of the Bloch sphere (Fig.~\ref{Fig:sequence}(b)). The population of $|1\rangle$ is measured by a probe detuned by $\Delta$ from the cyclic $|1\rangle - |3\rangle$ transition. Note that this probe does not cause redistribution of the populations between the clock states, hence $J_z$ is conserved and is a true QND variable~\cite{appel2009, RMP,Chen2014}. After a $\pi$-pulse is applied to swap the clock states, the population of $|1\rangle$ is measured again. Under this operation the effect of the imprecision of the $\pi/2$ pulse and fluctuations of $N$ are suppressed. This QND probe will introduce a Stark shift of the clock transition consisting of a mean value and a random shift due to the shot noise of the probe (quantum back action of the measurement). The former can be canceled by choosing the detuning $-\Delta$ for the second measurement of the population. We note that the precision of the $\pi$ rotation should be better than $N^{-1/2}$. This sequence results in creation of an SSS shown as an ellipse in Fig.~\ref{Fig:sequence}(c). Squeezing of $J_z$ is then converted into squeezing of the coherence between the clock states through a $\pi/2$ rotation around the $x$-axis (Fig.~\ref{Fig:sequence}(d)). The atomic spin is then let to precess, as in a standard Ramsey sequence (Fig.~\ref{Fig:sequence}(e)). After a certain precession time, a $\pi/2$ rotation around $y$ is applied (Fig.~\ref{Fig:sequence}(f)), where the population measurement noise is reduced by squeezing.

\begin{figure}
\center\includegraphics[width=1\columnwidth]{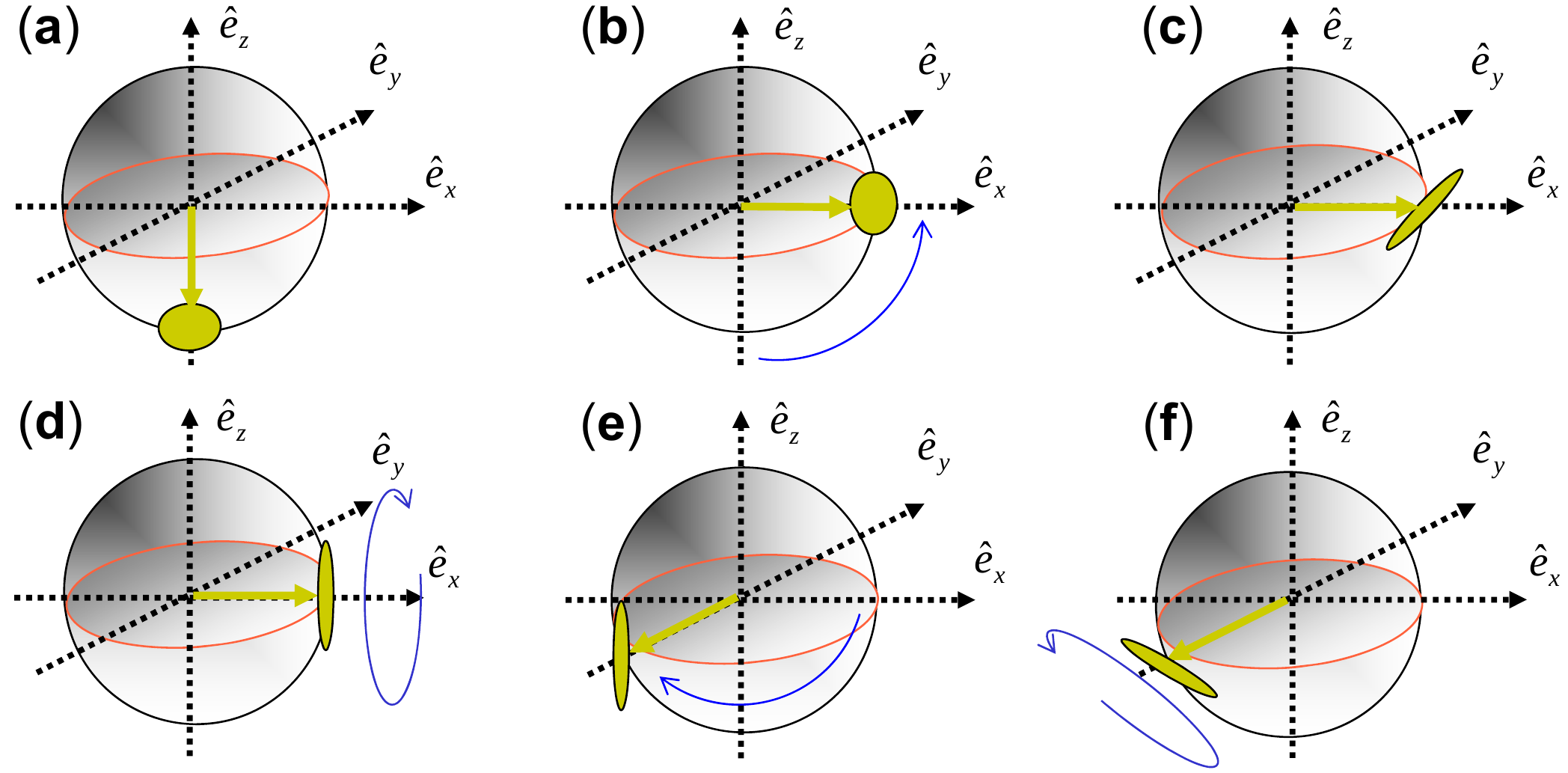}
\caption{\textbf{Entangled clock sequence}. The sequence of operations of the clock including generation of a spin squeezed entangled state. Details in the text. }\label{Fig:sequence}
\end{figure}

\noindent\textbf{Heisenberg scaling for optical clock precision with the number of atoms.} A coherent spin state (CSS) is a product state of individual uncorrelated atoms oriented in the same direction, for example, $J_x=N/2$, with $|\Psi\rangle = \Pi_{i=1}^{N} \frac{1}{\sqrt{2}}
(|1\rangle_i+|2\rangle_i)$. The other two projections of $\hat{J}$ have minimal equal variances allowed by the
Heisenberg uncertainty relation: $Var(J_z)=Var(J_y) = J_x/2 = N/4$. These fluctuations, referred to as QPN, and shown as a circle in Fig.~\ref{Fig:sequence}(b), pose a fundamental limit to the precision of the clock operating on $N$ uncorrelated atoms~\cite{itano1993}. Introducing quantum correlations between atoms allows to reduce $Var(J_z)$ to below the QPN limit. For this SSS, under the condition that
\begin{align} Var(J_z) {}& < \frac{J^2}{N} \nonumber\\
  \Leftrightarrow \xi = {}&
  \frac{Var(J_z)}{J^2}N =Var(X_A)\frac{N}{J}<1
  \label{eq:SMwineland},
\end{align}
the atoms become entangled~\cite{sorensenduan2001}, and the corresponding signal-to-noise ratio for spectroscopy is improved by the inverse of $\xi$ ~\cite{wineland1992}, which is the spin squeezing parameter for metrology. Here we introduce the canonical operators, $X_A=J_z/\sqrt{J},P_A=J_y/\sqrt{J}$, obeying the commutation relation $[X_A,P_A]=i$  where $J$ is the length of the mean pseudo-spin vector, which is the atomic coherence. The quantum noise limited clock precision defined as the minimal detectable angle of spin rotation for the clock sequence is $\sqrt{Var(J_z)}/J=\sqrt{\xi/N}$.

Eq.~\ref{eq:SMwineland} shows that $\xi$  is determined by the variance of the squeezed component of the spin and by the mean spin $J$.
 The input-output relation for light, with similarly defined operators of $X_L$ and $P_L$, assuming the QND interaction $H \propto X_A X_L$ ~\cite{RMP} is $P^{out}_L=P^{in}_L+\kappa X_A$.
To within a factor of unity the interaction constant $\kappa = \sqrt{d\eta e^{-\eta}}$, and $d$ is the resonant single pass optical depth. Furthermore, $\eta=n_{dr}(\gamma/\Delta)^2\sigma/A$ is a parameter describing spontaneous emission caused by the probe, which leads to the reduction of coherence as $J=e^{-\eta}N/2$. $\sigma$ is the resonant dipole cross section, $A$ is the beam cross section, $\gamma,\Delta$ the natural linewidth and detuning of the optical transition, $n_{dr}$ is the photon number in the QND probe. Taking the reduction of $J$ into account, we get the SSS with  $\xi=\frac{1}{e^{-\eta}\, (1+\kappa^2)}$. 
 The minimal value $\xi_{min}=\sqrt{e}/{\kappa_{opt}^2}= 2e/d\propto N^{-1}$ is achieved for $\eta=1/2$, and the optimal $\kappa_{opt}=\sqrt{d/2}$ valid for $d>>1$. The precision of the clock $\sqrt{\xi/N}$ then follows the Heisenberg scaling $1/N$. 

\begin{thebibliography}{clock}

\bibitem{Bloom2014} B. J. Bloom \emph{et al}, \emph{Nature} \textbf{506}, 71-75 (2014).

\bibitem{Nicholson2015} T. L. Nicholson \emph{et al}, \emph{Nature Comm.} \textbf{6}, 6896/1-8 (2015).

\bibitem{Katori2015} I. Ushijima \emph{et al}, \emph{Nature Photonics} \textbf{9}, 185 (2015).

\bibitem{Chou2010} C. W. Chou \emph{et al}, \emph{Phys. Rev. Lett} \textbf{104}, 070802 (2010).

\bibitem{Madej2012} A. A. Madej \emph{et al}, \emph{Phys. Rev. Lett} \textbf{109}, 203002 (2012).

\bibitem{Hinkley2013} N. Hinkley \emph{et al}, \emph{Science} \textbf{341}, 1215-1218 (2013).

\bibitem{Bishof2013} M. Bishof \emph{et al}, \emph{Phys. Rev. Lett.} \textbf{111}, 093604 (2013).

\bibitem{Kessler2012} T. Kessler \emph{et al}, \emph{Nature Photonics} \textbf{6}, 687-692 (2012).

\bibitem{Borregaard2013} J. Borregaard and A. S. S{\o}rensen, \emph{Phys. Rev. Lett.} \textbf{111}, 090802 (2013).

\bibitem{Rosenband2013} T. Rosenband and D. R. Leibrandt, arXiv:1303.6357 (2013).

\bibitem{Kohlhaas2015} R.Kohlhaas \emph{et al}, \emph{Phys. Rev. X} \textbf{5}, 021011 (2015).

\bibitem{Kessler2014} E. M. Kessler \emph{et al}, \emph{Phys. Rev. Lett.} \textbf{112}, 190403 (2014).

\bibitem{Komar2014} P. K\'{o}m\'{a}r \emph{et al}, \emph{Nature Phys.} \textbf{10}, 582 – 587 (2014).

\bibitem{wineland1992} D. J. Wineland \emph{et al}, \emph{Phys. Rev. A} \textbf{46}, R6797 (1992).

\bibitem{hald1999} J. Hald \emph{et al}, \emph{Phys. Rev. Lett.} \textbf{83}, 1319 (1999).

\bibitem{kuzmich2000} A. Kuzmich, L. Mandel, and N. P. Bigelow, \emph{Phys. Rev. Lett.} \textbf{85}, 1594 (2000).

\bibitem{julsgaard2001} B. Julsgaard, A. Kozhekin, and E. S. Polzik, \emph{Nature} \textbf{413}, 400 (2001).

\bibitem{appel2009} J. Appel \emph{et al}, \emph{Proc. Nat. Acad. Sci.} \textbf{27}, 10960 (2009).

\bibitem{schleier2010} M. H. Schleier-Smith, I. D. Leroux, and V.
  Vuletic, \emph{Phys. Rev. Lett.} \textbf{104}, 073604 (2010).

\bibitem{Leroux2010} I. D. Leroux, M. H. Schleier-Smith, and V. Vuletic, \emph{Phys. Rev. Lett.} \textbf{104}, 250801 (2010).

\bibitem{takano2009} T. Takano \emph{et al}, \emph{Phys. Rev. Lett.} \textbf{102}, 033601 (2009).

\bibitem{thompson2014} J. G. Bohnet \emph{et al}, \emph{Nat. Photonics} \textbf{8}, 731-736 (2014).

\bibitem{appel2008} J. Appel \emph{et al}, \emph{Phys. Rev. Lett.} \textbf{100}, 093602 (2008); K.
  Honda \emph{et al}, \emph{Phys. Rev. Lett.} \textbf{100}, 093601 (2008).

\bibitem{esteve2008} J. Est\`eve \emph{et al}, \emph{Nature} \textbf{455}, 1216 (2008).

\bibitem{krauter2011} H. Krauter \emph{et al}, \emph{Phys. Rev. Lett.} \textbf{107}, 080503 (2011).

\bibitem{Saffman2009} M. Saffman, D. Oblak, J. Appel, and E. S. Polzik, Phys. Rev. A \textbf{79}, 023831 (2009).

\bibitem{itano1993} W. M. Itano \emph{et al}, Phys. Rev. A \textbf{47}, 3554 (1993).

\bibitem{sorensenduan2001} A. S\o rensen, L. M. Duan, J. I. Cirac, and
  P. Zoller, Nature \textbf{409}, 63 (2001).

\bibitem{RMP} A. S\o rensen, K. Hammerer, E.S. Polzik, \emph{Rev. Mod. Phys.} \textbf{82}, 1041–1093 (2010).

\bibitem{anne} A. Louchet-Chauvet \emph{et al}, \emph{New J. Phys.} \textbf{12}, 065032 (2010).

\bibitem{Chen2014} Z. Chen \emph{et al}, \emph{Phys. Rev. A} \textbf{89}, 043837 (2014).

\bibitem{Bishof2014} M. Bishof, Ph.D. Thesis, Univ. of Colorado (2014).

\bibitem{Nicholson2012} T. Nicholson \emph{et al}, \emph{Phys. Rev. Lett.} \textbf{109}, 230801 (2012).

\bibitem{brukner} M. Zych \emph{et al}, \emph{Nat. Comm.} \textbf{2}, 605 (2011).

\bibitem{Bize2000} S. Bize, Y. Sortais, and P. Lemonde, \emph{ IEEE Trans. Ultrason. Ferroelectr. Freq. Ctrl.} \textbf{47}, 1253 (2000).

\bibitem{Takamoto2011} M. Takamoto, T. Takano, and H. Katori, \emph{Nat. Photo.} \textbf{5}, 288 (2011).

\bibitem{julsgaard} B. Julsgaard, A. Kozhekin, and E. S. Polzik, \emph{Nature} \textbf{413}, 400 (2001).

\bibitem{wasilewski2009} W. Wasilewski \emph{et al}, \emph{Phys. Rev. Lett.} \textbf{104}, 133601 (2010).

\bibitem{pikovski} I. Pikovski \emph{et al}, \emph{Nat. Phys.} DOI:10.1038/NPHYS3366.

\bibitem{Hogan2015} J. M. Hogan and M. A. Kasevich, arXiv:1501.06797 (2015).

\bibitem{Lodewyck2009} J. Lodewyck, P. G. Westergaard, and P. Lemonde, \emph{Phys. Rev. A} \textbf{79}, 061401 (2009).

\end{thebibliography}
\end{document}